\documentclass{iaus}
\usepackage{graphicx}
  \newcommand{\Msun}{M$_{\odot}$}

\newcommand{\Ha}{H$\alpha$}

\newcommand{\J}{M$_{JUP}$}
\newcommand{\ISO}{ISO-ChaI 217 }
\newcommand{\MASS}{2MASS1207-3932}
 \newcommand{\apj}{Astrophysical Journal}
 \newcommand{\apjl}{Astrophysical Journal Letters}
\newcommand{\aap}{Astronomy $\&$ Astrophysics}
\newcommand{\araa}{Astronomy $\&$ Astrophysics Review}
\newcommand{\nat}{Nature}

\title[Brown Dwarf Jets: Investigating the Universality of Jet Launching Mechanisms at the Lowest Masses] 
{Brown Dwarf Jets: Investigating the Universality of Jet Launching Mechanisms at the Lowest Masses}

\author[E.T. Whelan]   
{Emma Teresa Whelan$^1$
 \and Francesca Bacciotti$^2$ \and Tom Ray$^{3}$ \and Catherine Dougados$^{1}$ }

\affiliation{$^1$Laboratoire d'Astrophysique de Grenoble, UMR 5571, BP 53, 38041 Grenoble Cedex 09, France \\ email: {\tt whelane@obs.ujf-grenoble.fr} \\[\affilskip]
$^2$ INAF-Osservatorio Astrofisico di Arcetri, Largo E. Fermi 5, 50125 Firenze, Italy, \\ Box
515, SE-75120 Uppsala, Sweden \\email: {\tt fran@arcetri.astro.it}\\[\affilskip]
$^3$Dublin Institute for Advanced Studies, Ireland \\email:{\tt tr@cp.dias.ie}}

\pubyear{2010}
\volume{275}  
\pagerange{}
\jname{Brown Dwarf Jets: Investigating the Universality of Jet Launching Mechanisms at the Lowest Masses}
\editors{Gustavo E. Romero \and Rashid A. Sunyaev \and Tomaso Belloni., eds.}
\begin{document}

\maketitle

\begin{abstract}
Recently it has become apparent that proto-stellar-like outflow activity extends to the brown dwarf (BD) mass regime. While the presence of accretion appears to be the common ingredient in all objects known to drive jets fundamental questions remain unanswered. The more prominent being the exact mechanism by which jets are launched, and whether this mechanism remains universal among such a diversity of sources and scales.  To address these questions we have been investigating outflow activity in a sample of protostellar objects that differ considerably in mass and mass accretion rate. {\it Central to this is our study of brown dwarf jets.} To date Classical T Tauri stars (CTTS) have offered us the best touchstone for decoding the launching mechanism. Here
we shall summarise what is understood so far of BD jets and the important constraints observations can place on models. We will focus on the comparison between jets driven by objects with central mass $<$ 0.1\Msun\ and those driven by CTTSs. In particular we wish to understand how the the ratio of the mass outflow to accretion rate compares to what has been measured for CTTSs.
\end{abstract}

\firstsection 

\section{Brown Dwarf Outflows}

Jets from CTTSs are traditionally probed at forbidden emission line (FEL) wavelengths and long-slit and integral field spectroscopic techniques have been hugely important in their study [\cite{Dougados00}]. That CTT-like outflows could be launched by actively accreting BDs was first suggested when high quality spectra revealed the presence of FEL regions in the optical spectra of BDs known to be accretors [\cite{Fernandez01}]. While the FEL regions were easily identified they were considerably fainter than those detected in the spectra of CTTSs and any extension in the form of an outflow was not apparent [\cite{Fernandez01}]. Hence their origin in an outflow could not be established. As the most intense forbidden emission coincides with the critical density region it is currently challenging to directly resolve a BD outflow in FELs and this goal has not yet been achieved. Our approach to this problem has been to obtain high quality spectra using the UV-Visual Echelle Spectrometer (UVES) on the European Southern Observatory's (ESO) Very Large Telescope (VLT) and to recover the spatial offset in the region of forbidden emission using spectro-astrometry [\cite{Whelan08}]. 

To date we have observed 5 optical outflows driven by BDs (Table 1). While overall these jets are T Tauri-like in nature each object has its own unique properties. ISO-Oph 102 was the first BD confirmed to be driving a jet. Interesting follow-up observations with the SMA conducted by \cite{Phan08} also detected a molecular outflow driven by this BD. The properties of the outflow agreed with what was observed in the optical by \cite{Whelan05}. This is only one of a small number of objects (including protostellar objects) known to be driving both an optical jet and a molecular outflow. 2MASS1207-3932 at only 24M$_{JUP}$ is the lowest mass galactic object known to drive a jet. The detection of a jet was first reported by \cite{Whelan07} and follow-up observations have constrained the position angle (PA) of the jet at $\sim$ 220$^{\circ}$. 2MASS1207-3932 is an intriguing object as it has a planetary mass companion which was imaged with the VLT [\cite{Chauvin05}]. Our recent observations show the jet PA to be approximately perpendicular to the PA of the companion suggesting the presence of circumbinary structure [Whelan et al. 2010, in prep]. This observation is relevant to the formation of the companion. In the case of LS-RCr A1 spectro-astrometry confirmed that both the line-wings of the H$\alpha$ line and the forbidden emission originated in a jet. What is interesting about this object is that while only blue-shifted forbidden emission is observed both lobes of the outflow are detected in the H$\alpha$  line. This is taken as evidence of a dust hole in the disk of LS-RCr A1 which suggests the onset of planet-forming processes [\cite{Whelan09a}]. Finally we will mention the jet driven by ISO-ChaI 217. We detected a bipolar outflow from this object however what was notable was the strong asymmetry between the two lobes of the outflow. This asymmetry is revealed in the relative brightness of the two lobes (red-shifted lobe is brighter), the factor of two difference in radial velocity (the red-shifted lobe is faster) and the difference in the electron density (again higher in the red lobe). Such asymmetries are common in jets from low mass protostars and the observation of a marked asymmetry at such a low mass ($<$ 0.1\Msun) supports the idea that BD outflow activity is scaled down from low mass protostellar activity. Also note that although asymmetries are unexceptional, it is uncommon for the red-shifted lobe to be the brightest as some obscuration by the accretion disk is assumed.


\section{Constraining Models}
Studies of BD outflows can offer constraints to both models of jet launching and BD formation. Current models describing the launching and collimation of protostellar jets predict how various jet parameters, including the ratio of the mass outflow to accretion rate, scale with mass and mass accretion rate ($\dot{M}_{out}$/$\dot{M}_{acc}$). Thus we have identified $\dot{M}_{out}$/$\dot{M}_{acc}$ as an important observational constraint and have thus begun the work of measuring this for sub-stellar outflows. To date we have estimated $\dot{M}_{out}$/$\dot{M}_{acc}$ for three of the known BD outflows. This was derived using two methods. Method A is based on the equation 

\begin{equation}
 \dot{M}_{out} =  \mu m_{H} n_{H} \pi r_{J}^{2} v_{J}
 \end{equation}

where n$_{e}$ , x$_{e}$ and thus n$_{H}$ are derived using the Bacciotti $\&$ Eisloffel (BE) technique (Bacciotti et al 1999). Reasonable estimates of the jet velocity v$_{J}$ and jet radius r$_{J}$ come from measured radial velocities and a previously derived relation between jet width and distance. Method B is based on the observed luminosity {\it L(line)} of an optically thin line such as [SII] or [OI]. The mass of the flow {\it M} is derived using the equations of Hartigan et al 1995.  $\dot{M}_{out}$ is estimated as follows, $\dot{M}_{out}$ = MV$_{tan}$/l$_{tan}$ where n$_{c}$, V$_{tan}$ and l$_{tan}$ are the critical density, outflow tangential velocity and the size of the aperture in the plane of the sky. Table 2 compares the values derived for $\dot{M}_{out}$ using both methods with previously known estimates of $\dot{M}_{acc}$. 
What is clear from these results is that can be stated at present is that the mass outflow and mass accretion rates are comparable. To further constrain their ratio the number of substellar outflows investigated must be greatly increased. It is also important to simultaneously derive the two rates and to better understand the sources of error in the various methods for estimating the mass accretion rate. See Whelan et al 2009(b) for further discussion of the methods used and the results. We also plan to investigate the occurrence of episodic jets in the BD regime and the frequency of molecular outflows. For low mass stars episodic jets are taken as evidence of variable accretion. If BDs jets are also found to episodic this will help to drive models of BD accretion activity. Projects to increase the sample of BD jets, further investigate $\dot{M}_{out}$/$\dot{M}_{acc}$ and search for BD molecular outflows are currently underway.

 \begin{table}
\begin{tabular}{lllll}       
 \hline\hline 
 Source                                   &RA (J2000)   &Dec (J2000) & Spectral Type &Mass (\J)     
 \\ 
\hline
ISO-ChaI 217          &11 09 52.0                  &-76 39 12.0  &M6.2                 &80$^{1}$             
\\
\MASS &12 07 33.4 &-39 32 54.0 &M8 &24$^{2}$
\\
DENIS-P J160603.9-205644  &16 06 03.90 &-20 56 44.6 &M7.5 & 40$^{3}$
\\
ISO-Oph 32         &16 26 22.05  &-24 44 37.5 &M8 &40$^{4}$  
 \\
 ISO-Oph 102      &16 27 06.58   &-24 41 47.9  &M6 &60$^{4}$
 \\
LS-RCr A1   &19 01 33.7  &-37 00 30.0 &M6.5 &35-72$^{5}$
\\  
 \hline  
\end{tabular}
\caption{The spectral type and predicted mass of the BD candidates investigated by us to date. All sources except DENIS-P J160603.9-205644 are found to drive outflows. The numbers 1-6 refer to the papers giving the estimates of the mass of each source, where 1=\cite{Muz05}, 2=\cite{Mohanty07}, 3=\cite{Mohanty04}, 4=\cite{Natta02} and 5=\cite{Barrado04}.
}
\label{tab1}
\end{table}

 \begin{table}
\begin{tabular}{ccc}       
 \hline\hline 
 Object                                 &$\dot{M}_{out}$ (\Msun yr$^{-1}$)    & Method 
 \\
 \hline
LS-RCr A1                                                  & & 
\\
                                                      &2.4 $\times$ 10$^{-9}$       &A                                          
\\
                                                      &6.1 $\times$ 10$^{-10}$     &B [OI]$\lambda$6300       
\\
                                                      &2.0 $\times$ 10$^{-10}$     &B [SII]$\lambda$6731                                                 
\\
 ISO-Oph 102                         & &                     
 \\
                                                       &1.7-11.8 $\times$ 10$^{-10}$        &B [SII]$\lambda$6731        
\\
                                                       &1.4 $\times$ 10$^{-9}$                    &CO molecular outflow$^{1}$
\\                                                       
\ISO Red flow
\\
                                                      &3.1 $\times$ 10$^{-10}$   &B [SII]$\lambda$6731
\\
\ISO Blue flow 
\\                                                     
                                             &1.8 $\times$ 10$^{-10}$   &B [SII]$\lambda$6731
\\
\hline

\end{tabular}

\vspace{1cm}

\begin{tabular}{ccc}       
 \hline\hline 
 Object                                 &$\dot{M}_{acc}$ (\Msun yr$^{-1}$)    & Method 
 \\
 \hline
LS-RCr A1                                                  & & 
\\
                                                      &2.8 $\times$ 10$^{-10}$       &CaII($\lambda$8542)$^{2}$                                         
\\
                                                      &10$^{-10}$-10$^{-9}$           &Optical Veiling$^{3}$     
\\
                                                      &10$^{-9}$                                &CaII($\lambda$8662)$^{4}$                                                  
\\
                                                      &10$^{-10}$                              &\Ha\ 10$\%$ width$^{5}$
\\  
 ISO-Oph 102                         & &                     
 \\
                                                       &10$^{-9}$        &\Ha\ 10$\%$ width$^{6}$         
\\
                                                       &4.3 $\times$ 10$^{-10}$  & J and K band spectra$^{7}$
                                                       \\                                                       
\ISO
\\
                                                      &1.0 $\times$ 10$^{-10}$ &\Ha\ emission$^{8}$ 
\\
 \hline

\end{tabular}
\label{tab6}
\caption{Measurements of $\dot{M}_{out}$ and $\dot{M}_{acc}$ for the objects studied to date. For the BDs in our sample $\dot{M}_{out}$ is comparable to $\dot{M}_{acc}$. 1=\cite{Phan08}, 2=\cite{Comeron03}, 3=\cite{Barrado04}, 4=\cite{Mohanty05}, 5=\cite{Scholz06}, 6=\cite{Natta04}, 7=\cite{Natta06}, 8=\cite{Muz05}}
\end{table}

\end{document}